\documentclass[aps,prx,twocolumn,superscriptaddress,longbibliography, nofootinbib]{revtex4-2}
\usepackage[T1]{fontenc}
\usepackage{color}
\usepackage{xcolor}
\usepackage{graphicx}
\usepackage[english]{babel}
\usepackage{amsmath}
\usepackage{dsfont}
\usepackage{pgfplots}
\usepackage{physics}
\usepackage{tikz}
\usepackage[unicode=true,pdfusetitle,
bookmarks=true,bookmarksnumbered=false,bookmarksopen=false,
 breaklinks=true,pdfborder={0 0 0},pdfborderstyle={},backref=false,colorlinks=true]
 {hyperref}
\usepackage[nolist,nohyperlinks]{acronym}
\usepackage{xspace}
\usepackage{subcaption}

\newcommand{\eviden}{Eviden Quantum Lab, Les Clayes-sous-Bois, France}
\newcommand{\mpdo}{\ac{MPDO}\xspace} 
\newcommand{\mpdos}{\acp{MPDO}\xspace}
\newcommand{\Mpdos}{\Acp{MPDO}\xspace}
\newcommand{\ourmethod}{\ac{IPD} technique\xspace}
\newcommand{\theirmethod}{\ac{LC} technique\xspace}
\newcommand{\theirmethodf}{\acf{LC} technique\xspace}

\newacro{NISQ}{noisy intermediate scale quantum}
\newacro{TN}{tensor network}
\newacro{MPS}{matrix product state}
\newacro{MPO}{matrix product operator}
\newacro{OEE}{operator entanglement entropy}
\newacro{MPDO}{matrix product density operator}
\newacro{LPDO}{locally purified density operator}
\newacro{SVD}{singular value decomposition}
\newacro{IPD}{iterative purification disentanglement}
\newacro{LC}{local compression}
\newacro{DMRG}{density matrix renormalization group}
\newacro{DMT}{density matrix truncation}
\newacro{LPF}{locally purified form}
\newacro{LPTN}{locally purified tensor network}
\newacro{MMS}{maximally mixed state}

\begin{document}

\title{Enabling large-depth simulation of noisy quantum circuits with positive tensor networks}

\author{Ambroise Müller}
\affiliation{\eviden}
\author{Thomas Ayral}
\affiliation{\eviden}
\author{Corentin Bertrand}
\email[]{corentin.bertrand@eviden.com}
\affiliation{\eviden}

\begin{abstract}
\Mpdos are tensor network representations of locally purified density matrices where each physical degree of freedom is associated to an environment degree of freedom.
\Mpdos have interesting properties for mixed state representations: guaranteed positivity by construction, efficient conservation of the trace and computation of local observables.
However, they have been challenging to use for noisy quantum circuit simulation, as the application of noise increases the dimension of the environment Hilbert space, leading to an exponential growth of bond dimensions.
\Mpdos also lack a unique canonical form, due to the freedom in the choice of basis for the environment Hilbert space, which leads to a vast variation of bond dimensions.

In this work, we present a systematic way to reduce the bond dimensions of \mpdos by disentangling the purified state.
We optimize the basis for the environment Hilbert space by performing \ac{DMRG}-like sweeps of local 2-qubit basis optimization.
Interestingly, we find that targeting only the disentanglement of the purified state leads to a reduction of the environment dimension.
In other words, a compact \mpdo representation requires a low-entanglement purified state.

We apply our compression method to the emulation of noisy random quantum circuits. Our technique allows us to keep bounded bond dimensions, and thus bounded memory, contrary to previous works on \mpdos, while keeping reasonable truncation fidelities.
\end{abstract}
\maketitle

\acresetall % reset acronyms

\section{Introduction}

With the progress of experimental platforms for quantum computing, it is becoming increasingly critical to produce accurate emulations of these machines on classical hardware, taking into account their imperfections.
Indeed, finding algorithms relevant for \ac{NISQ} hardware and appropriate noise-mitigation techniques \cite{Cai2022} requires an understanding of the effect of noise.
The controlled environment of a classical emulator for large noisy quantum circuits is therefore a precious tool.

Besides perfect (compressionless) emulators, whose memory and run time scale exponentially in the number of qubits or gates (or both),
approximate emulators using the fruitful concept of \acp{TN} have emerged in the last decade as methods of choice for this task~\cite{ayral2023, tindall2023}.
The simplest form of \ac{TN}, a one-dimensional network called the \ac{MPS}, allows for good approximations of weakly entangled pure states~\cite{paeckel2019, cirac2021}, and for the emulation of noiseless quantum circuits with hundreds of qubits~\cite{vidal2003, banuls2006, dang2019, ayral2023}.
Other types of \acp{TN} better suited to entanglement patterns in higher than 1D geometries have been used, such as tree tensor networks~\cite{kloss2020a}, multi-scale entanglement renormalization ansatz~\cite{evenbly2014}, projected entangled pair states~\cite{cirac2021}, or more recently isometric tensor network states~\cite{zaletel2020, lin2022}.

In noisy quantum computing, the state is represented by a density operator $\rho$, as opposed to a pure state $\ket{\Psi}$, which a priori leads to an increased memory footprint.
Yet, \ac{TN} techniques can be applied to density operators as well, with the intuition that more noise will lead to less entanglement and therefore more representative power of compressed tensor networks \cite{noh2020, Oh2021}.
A natural extension of \ac{MPS} to density operators are \acp{MPO}~\cite{Verstraete2004, zwolak_mixed_state_2004, Prosen2009, GuthJarkovsky2020, noh2020, Oh2021, DeLasCuevas2020}: a product of matrices, each of them acting locally on a single (or a few) qubits.
Despite being powerful compression tools for mixed states, \acp{MPO} suffer from some important flaws.
First, the positive-semidefinite character of the density matrix is not easily enforced in \ac{MPO}, and may be lost in particular during the truncation of bond dimensions, leading to unphysical states.
Second, taking expectation values and normalizing the state require global operations, whereas \ac{MPS} can be cast in a canonical form that allows for local operations only.
Finally, the approximation made with an \ac{MPO} is controlled by the \ac{OEE}~\cite{rath2023}---instead of the entanglement entropy for pure states---
whose relationship with entanglement is not straightforward. 

There exists two alternatives to \acp{MPO}, both rooted in the idea of coming back to pure states.
The first is to unravel the density operator into a statistical population of pure states of the same Hilbert space, leading to powerful Monte Carlo methods~\cite{jaschke2018}.
The second, which is the focus of this work, is to purify the density operator in a larger Hilbert space, extending each original qubit with an ``environment'' or ``ancilla'' qudit of dimension $r$.
This purified state can be compressed into an \ac{MPS}, with an internal bond dimension $\chi$.
This gives a locally purified \ac{TN} representation with a memory footprint proportional to $r\chi^2$, that we call in this work \mpdo\footnote{\mpdos are sometimes called \acp{LPTN}, \acp{LPDO} or \acp{LPF}. Also, \acp{MPO} are sometimes called MPDOs. See App.~\ref{app:terminology}.}~\cite{Verstraete2004, cuevas2013, werner2016, jaschke2018, DeLasCuevas2020}.
This form is guaranteed to be positive semi-definite, and it is guaranteed to represent a physical state.
Moreover, in its canonical form, expectation values and normalization can be obtained with local operations only.
\Mpdos have been mostly used in quantum many-body physics, e.g. for the unitary evolution of mixed states \cite{hauschild2018, pokart2023}, imaginary-time evolutions \cite{jaschke2018, hauschild2018}, or the evolution of open quantum systems under a Lindblad equation \cite{werner2016, jaschke2018}.
In Ref.~\cite{surace2019}, \mpdos are used as an approximation for pure states, conserving local correlations only. 
In quantum computing \mpdos have also been used to perform tomography \cite{guo_scalable_2023}.

Nevertheless, \mpdos have been challenging to use for the simulation of noisy quantum circuits.
Similarly as with \acp{MPO}, entangling gates cause an increase of the bond dimension $\chi$, which needs to be reduced to avoid an exponential growth in the number of entangling layers.
A first reduction of $\chi$ can be obtained with routine \ac{MPS} compression techniques based on the \ac{SVD}~\cite{schollwock_matrix_2013}.
In addition, the freedom in the choice of basis for the environment can be beneficial in reducing the entanglement of the purified state.
Yet, a global optimization is required to take full benefit of this freedom, and can be done with successions of 2-sites local optimizations \cite{hauschild2018, Nguyen2018}.
Other ideas were also studied, such as Uhlmann parallel transport \cite{pokart2023}, or the application of backward time evolution to the environment \cite{karrasch_finite-temperature_2012, karrasch_reducing_2013, barthel_precise_2013}.
For \acp{MPO}, different techniques have also been developed, such as \ac{DMT}~\cite{white2017}.

On top of the increase in $\chi$, however, applying a noisy gate to an \mpdo increases the environment dimension $r$, leading to an exponential growth as a function of the number of noisy layers, a serious drawback compared to \acp{MPO}.
Note that this issue is absent from real or imaginary-time evolution of closed systems.
A straightforward idea to try to reduce $r$ is to perform truncated \ac{SVD} on the environment bonds, which boils down to an optimization of single-site environment bases followed by a truncation of unused dimensions \cite{werner2016, jaschke2018, cheng2021}.
Nevertheless, this \theirmethod, due to its single-site nature, does not avoid an exponential increase of $r$, as can be seen from Ref.~\cite{cheng2021}.
Extending it to a global optimization of the environment basis seems a natural way to go, but raises some questions.
While the optimization problem to reduce $\chi$ has a clear cost function, i.e. the entanglement entropy of the purified state, here we face the coupled problem of reducing both $\chi$ and $r$.
From an optimization point of view, there is no obvious choice of cost function.
Note that ideas that do not involve the environment basis freedom have also been explored, such as modifying the environment sites distribution along the \mpdo, e.g. to gather all the environment dimensions on a single site~\cite{guo2022}.

In this work, we show that the global optimization of the environment basis targeting minimal entanglement of the purification leads to drastic reduction in $r$, and thus forms a good compression scheme for \mpdos.
Such an optimization is expected to reduce $\chi$ \cite{hauschild2018, Nguyen2018}, but its impact on $r$ was never thoroughly studied.
We perform a global change of the environment basis with \ac{DMRG}-like sweeps of local 2-qubit optimizations, each of which disentangles the two qubits.
We will refer to this procedure as the \ourmethod.
We find that aiming at disentangling the purified state is enough to reduce both $r$ and $\chi$, giving a simple cost function for the compression procedure.
In fact, we observe that we can systematically reduce $r$ to a value close to the optimal value of $2$, with a very high truncation fidelity.
Thanks to this compression, we are able to emulate a noisy quantum random circuit with \mpdos at arbitrary depth with bounded memory, which was impossible with the \theirmethod.

This article is structured as follows.
We start by a pedagogical introduction to \mpdos in Sec.~\ref{sec:mpdo_facts}, followed to a description of the compression method in Sec.~\ref{sec:compression_method}.
Results are presented in Sec.~\ref{sec:results}.
First, the compression of a simple two-qudit state is studied (\ref{sec:two_qudits}), before moving to a many-qubit state (\ref{sec:one_shot_compression}).
Application to a full circuit simulation is shown in Sec.~\ref{sec:application_to_circuits}.
Discussion and conclusion are regrouped in Sec.~\ref{sec:conclusion}.

\section{Method}\label{sec:method}

In this section we first give an overview of the representation of mixed quantum states with \ac{MPDO}s. We then detail the \ourmethod used to obtain compressed \mpdo representations. 

\subsection{\mpdo states: general considerations}
\label{sec:mpdo_facts}

\begin{figure}[t]
    \centering
    \includegraphics[width=0.36\textwidth]{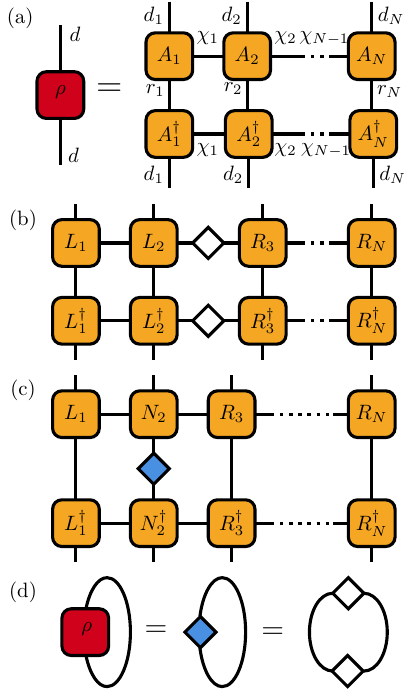}
    \caption{
    (a) Tensor network representation of a \mpdo state.
    (b) Extraction of singular values on entanglement bond at (bond) index $i=2$.
    (c) Extraction of singular values on purification bond at index $i=2$.
    (d) Normalization identities for singular value distributions.
    }
    \label{fig:mpdo}
\end{figure}

\begin{figure*}[t]
    \centering
    \includegraphics[width=0.92\textwidth]{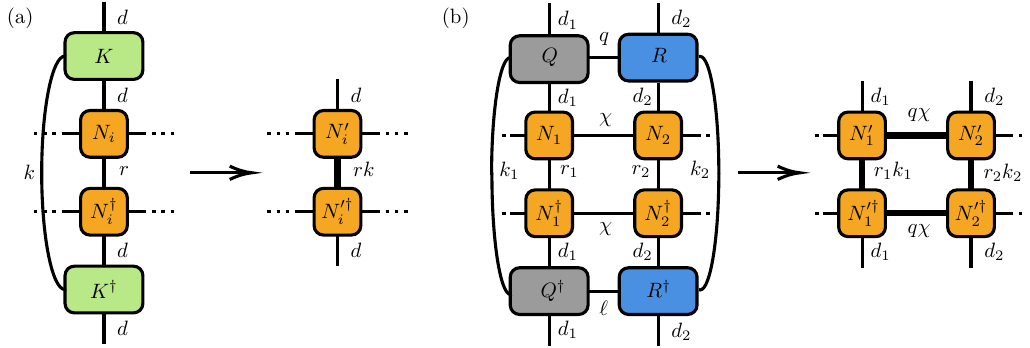}
    \caption{
    Tensor network representation of the application to a \mpdo state of (a) a one-site noisy channel with Kraus operators $\{K_i\}_{i=1}^k$, and (b) a two-site noisy channel.
    For the two-site case a QR decomposition is first performed on the two-qubit Kraus operators (allowing for an even splitting of the Kraus index) and the resulting tensors are then contracted with the site tensors.
    The purification bond dimensions are multiplied by the local Kraus ranks $k$, $k_1$, $k_2$, and the entanglement bond dimension is multiplied by the QR rank $q$.
    }
    \label{fig:noisy-channel}
\end{figure*}

Given a Hilbert space $\mathcal{H}$ and a mixed state $\rho$ on $\mathcal{H}$, a purification of $\rho$ is a pure state $\ket\Psi$ defined on an extended Hilbert space $\mathcal{H} \otimes \mathcal{H}'$, where $\mathcal{H}'$ is an ancillary system, and such that 
\begin{equation}\label{eq:purification}
    \rho = \text{Tr}_{\mathcal{H}'} \ket{\Psi} \bra{\Psi}.
\end{equation}
$\text{Tr}_{\mathcal{H}'}$ denotes the partial trace on the ancillary system.
Purifications always exist and are not unique in general.
The ancillary Hilbert space $\mathcal{H}'$ need not have the same dimension as $\mathcal{H}$, though its dimension must be at least equal to the rank of $\rho$. In particular there always exists a purification of rank $2^N$ for an arbitrary state of $N$ qubits.
As any pure state, a purification may be written in \ac{MPS} form:
\begin{equation}\label{eq:mps}
    \ket{\Psi} = \sum_{\{s_i\}} \sum_{\{a_i\}} \Tr{\prod_{i=1}^N A_i^{s_i a_i}} \ket{s_1 a_1, \dots \ , s_N a_N },
\end{equation}
where $s_i \in \{1, ..., d_i\}$ are the indices of physical degrees of freedom ($d_i \equiv 2$ for a system of qubits) and $a_i \in \{1, ..., r_i\}$ correspond to the ancillary degrees of freedom. For each $i$, $A_i^{s_i a_i}$ is a matrix of dimension $\chi_i \times \chi_{i+1}$.
We will refer to $\chi_i$ as \emph{entanglement bond dimensions} and to the corresponding bonds as \emph{entanglement bonds}. By contrast, we refer to $r_i$ as \emph{purification bond dimensions}, and to the corresponding bonds as \emph{purification bonds}.

We can combine Eqs.~\eqref{eq:purification} and \eqref{eq:mps} into the \ac{TN} diagrammatic expression of Fig.~\ref{fig:mpdo}a.
The physical Hilbert space has dimension $d = \prod_{i=1}^N d_i$ and the ancillary system has dimension $r = \prod_{i=1}^N r_i$.
The physical state is obtained by tracing out the purified state over the ancillary degrees of freedom, resulting in internal bonds.
Note that since the tensors in the bottom row are adjoints to those in the top row, one only needs to store one of them (i.e. the purified state vector or its dual) in memory. 

The \acf{SVD} is an essential tool in \ac{TN}-based methods.
It allows to associate a singular value distribution to any internal bond.
\ac{SVD} can be used to fix gauge freedom intrinsic to (non-looped) tensor networks and introduce canonical forms \cite{Bridgeman_2017}, yielding left-unitary and right-unitary tensors (which we denote by $L$ and $R$ respectively) and `center' the representation on a particular bond, of which singular values are extracted.
Figures~\ref{fig:mpdo}b and~\ref{fig:mpdo}c show the extraction of singular values across an entanglement bond and a purification bond, respectively, in the appropriately-centered canonical forms.
The singular value distributions associated with entanglement bonds are represented by the white diamonds, and will be denoted $\qty{\sigma_j^{(i)}}_{j=1}^{\chi_i}$ for bond index $i$.
On the other hand, the singular value distributions associated with purification bonds are shown as a blue diamond, and will be denoted $\qty{\lambda_j^{(i)}}_{j=1}^{r_i}$ for bond index $i$.

These singular values satisfy the normalization conditions shown in diagrammatic notation in Fig.~\ref{fig:mpdo}d and which reads
\begin{equation}
    \Tr\rho = \sum_{j=1}^{r_i} \lambda_j^{(i)} = \sum_{j=1}^{\chi_i} \qty(\sigma_j^{(i)})^2
\end{equation}
for all $i = 1, \ldots, N$.
For a normalized state ($\Tr\rho=1$) this enables us to associate a probability distribution to each \mpdo bond,
to which we can associate an entropy measure.
In this vein, if $X$ is a positive semidefinite matrix whose singular values $\qty{x_j}_{j=1}^n$ form a probability distribution, we define the R\'enyi entropies \cite{Renyi1961} 
\begin{equation}\label{renyi-entropies}
    E_\alpha \left[ X \right] = \frac{1}{1-\alpha} \log \left( \sum_{j=1}^n x_j^\alpha \right) ,
\end{equation}
where $\alpha \ge 0$ is the entropy order. The (common) choices $\alpha=1$ and $\alpha=2$ simplify to 
\begin{align}
    E_1[X] &= - \Tr{X \log X }, \label{eq:entropies1} \\
    E_2[X] &= - \log \Tr{X^2}. \label{eq:entropies2} 
\end{align}

We will refer to the entropy of the distribution associated to a bond as the \emph{entanglement bond entropy} or \emph{purification bond entropy}, denoted respectively by
\begin{align}\label{eq:bond-entropies}
    \mathcal{E}_{\alpha}^i = E_{\alpha}\left[ (\Sigma^{(i)})^2 \right] \ \text{and} \
    \mathcal{P}_{\alpha}^i = E_\alpha\left[ \Lambda^{(i)} \right],
\end{align}
where we defined the diagonal singular value matrices $\Sigma^{(i)} = \text{diag}\left( \qty{ \sigma_j^{(i)}}_{j=1}^{\chi_i} \right)$ and $\Lambda^{(i)} = \text{diag}\left( \qty{ \lambda_j^{(i)}}_{j=1}^{r_i} \right)$. 
% \am{maybe move definition earlier, with $\sigma, \lambda$}
Note that $\mathcal{E}_\alpha^i$ is the $\alpha$-entanglement entropy of the purified state between left and right of index $i$ (both sides including system and ancilla degrees of freedom). $\mathcal{E}_2^i$ is therefore related to the purity of the corresponding reduced density matrix. Moreover, $\mathcal{P}_\alpha^i$ is the $\alpha$-entanglement entropy between the ancilla degree of freedom at index $i$ and the all other degrees of freedom.

The simulation of noisy quantum circuits with \mpdos is rather straightforward once the circuit has been decomposed into one-qubit and nearest-neighbour two-qubit channels.
Distant two-qubit channels can be made nearest-neighbour by introducing a series of SWAP gates \cite{Li2019a,Martiel2020}, even though this incurs a large increase in the number of entangling gates.
A generic quantum channel is a completely positive trace-preserving map \cite{Renes_2022}, and admits the decomposition
\begin{equation}
    \rho \longrightarrow \sum_{\ell=1}^k K_\ell \rho K^\dagger_\ell ,
\end{equation}
where $\{K_\ell\}_{\ell=1}^k$ are known as Kraus operators.
The sum over the Kraus index (of dimension $k$) can be construed as an additional bond in \ac{TN} diagrammatic notation, as illustrated in Fig.~\ref{fig:noisy-channel}.
Applying the channel consists in contracting the Kraus operators with the qubit site tensors and grouping the Kraus index with the purification bond.
This has the result of increasing the purification bond dimension by a factor $k$.
In an exact circuit simulation, the purification bond dimensions thus grow exponentially with circuit depth.
Two-qubit gates act in a similar way, though also increasing the dimensions of entanglement bonds, much in the same way as with \ac{MPS} or \ac{MPO} simulations.

\subsection{\Acf{IPD}}
\label{sec:compression_method}

Our proposed method relies on the fact that purifications of a mixed quantum state are not unique.
For any mixed state one is free to construct a purification requiring an ancilla system much larger than the physical system, yielding large purification bond dimensions (in \mpdo form), or which are highly entangled even for a separable physical state, yielding large entanglement bond dimensions. 
In order to reach a memory-efficient representation, it is thus necessary to find a purification of small rank with minimal entanglement, i.e. with minimal quantum correlations between ancillary degrees of freedom but still retaining the quantum correlations in the physical system.
Given two purifications $\ket\Psi \in \mathcal{H}\otimes\mathcal{H}'$ and $\ket\Psi' \in \mathcal{H}\otimes\mathcal{H}''$ of a quantum state $\rho$ there exists a partial isometry $U:\mathcal{H}'\rightarrow \mathcal{H}''$ (i.e. $U^\dag U$ is a projector onto $\mathcal{H}'$, and, equivalently, $U U^\dag$ is a projector onto $\mathcal{H}''$) such that
\begin{equation}
    \ket\Psi' = (\mathds{1}_{\mathcal{H}} \otimes U) \ket\Psi .
\end{equation}
In particular, $U$ is unitary when $\dim\mathcal{H}' = \dim\mathcal{H}''$\cite{Renes_2022} and can be interpreted as a change of basis for the ancilla system. 
This translates to a gauge freedom in the \mpdo representation, whereby a unitary transformation may be applied across any number of purification bonds without altering the physical state.

Applying this gauge freedom with single-site operators $U$ allows to change the local basis of the ancilla system.
An optimal local basis to reduce the purification bond dimensions can be obtained with an \ac{SVD}, truncating the lowest singular values, as in Refs.~\cite{werner2016, cheng2021}.
We will refer to this as the \theirmethodf.

\begin{figure}[t]
    \centering
    \includegraphics[width=0.46\textwidth]{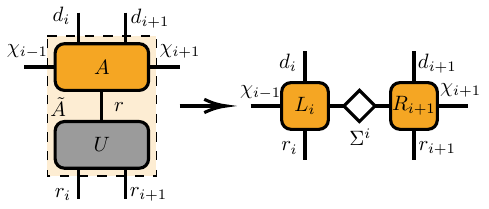}
    \caption{
    Upon contracting two adjacent sites of the purification MPS into a tensor $A$, one is free to apply a gauge unitary $U$ on the ancillary indices without affecting the physical state.
    The choice of unitary will affect the distribution of singular values $\Sigma^{(i)}$ across the entanglement bond upon re-separation of the sites via SVD of the transformed tensor $\tilde A = L \Sigma^{(i)} R$.
    }
    \label{fig:entglt_optim}
\end{figure}

Now, consider two adjacent sites, at indices $i$ and $i+1$. 
An optimal basis for these two sites cannot be found simply with an \ac{SVD}, as the size of the entanglement bond between the two sites must be taken into account.
Instead, we first contract the sites into a tensor $A$ and group the purification bonds into a single bond of dimension $r = r_i r_{i+1}$.
We apply an optimized $r \times r$ unitary $U$ on the ancilla index. 
As illustrated in Fig.~\ref{fig:entglt_optim}, this unitary can be understood as a bond partitioning operation, whereby the contracted local purification bond of dimension $r$ is factorized again into two bonds of dimension $r_i$ and $r_{i+1}$.  
Denoting by $\tilde A$ the gauge-transformed tensor,
where left and right indices are grouped to form a matrix of shape $d_i \chi_i r_i \times d_{i+1} \chi_{i+1} r_{i+1}$, one can extract the entanglement bond entropy $\mathcal{E}_\alpha^i = E_\alpha\qty[\tilde A^\dag \tilde A]$.

We choose to optimize $U$ so as to minimize this entropy.
This means that we need to solve the problem
\begin{equation}
\begin{aligned}\label{eq:entropy-opt}
    &\underset{U}{\text{minimize}} \ \mathcal{E}_\alpha^i \\ % E\left[ \{\sigma_j\}_j \right] \\
&\text{such that} \ UU^\dagger = U^\dagger U = \mathds{1}.
\end{aligned}
\end{equation}
Perhaps surprisingly, we choose a cost function that depends only on the entanglement bond, and no the purification bond.
This is motivated by the intuition that a purification with a simpler entanglement structure might be able to be described by fewer degrees of freedom.
We numerically check that both bond dimensions will be reduced with this cost function in Sec.~\ref{sec:results}.

Any minimum of Eq.~\eqref{eq:entropy-opt} is necessarily non-unique, since we can identify equivalence classes of gauge unitaries.
Indeed, two unitaries which are related through the tensor product of unitaries of shape $r_i\times r_i$ and $r_{i+1}\times r_{i+1}$ (acting on the separated purification bonds), i.e. where only the single-site gauge choice differs, will yield the same entanglement bond entropy.

The optimization problem in Eq. \eqref{eq:entropy-opt}, with either entropy measure, can be solved via gradient descent.
In our results we used the implementation of the conjugate gradient (CG) algorithm provided by SciPy.
Details of the gradient computation are shown in App.~\ref{app:optimization}.
An a priori advantage of $\mathcal{E}_2^i$ over $\mathcal{E}_1^i$, as already observed in Ref.~\cite{hauschild2018}, is the absence of a matrix logarithm, thus simplifying the function and gradient evaluations.

To generalize to $N$ qubits, we do a sequence of sweeps along the \mpdo, in alternating directions.
A sweep is performed in a two-site DMRG fashion, that is, over pairs of adjacent sites at indices $i$ and $i+1$, for $i = 1$ to $i = N-1$.
On each pair of sites, we solve problem~\eqref{eq:entropy-opt} and apply the corresponding disentangling operator $U$.
Upon separating the two sites via SVD, we usually see an increase in bond dimension, due to the introduction of small singular values.
To get the benefits of the change of representation, we truncate singular values that fall under a given threshold $\varepsilon$.
After every sweep, we also truncate singular values on local purification bonds below this threshold.
A small $\varepsilon$ guarantees a small truncation fidelity after a set number of sweeps, but a low compression rate.
A high $\varepsilon$ gives rise to a better compression rate, but at the risk of a reduced fidelity.
Truncation error is guaranteed to be bounded by the total weight of removed singular values, for both the entanglement and purification bond, thanks to the canonical form.

\begin{figure*}[t]
    \centering
    \includegraphics[width=0.96\textwidth]{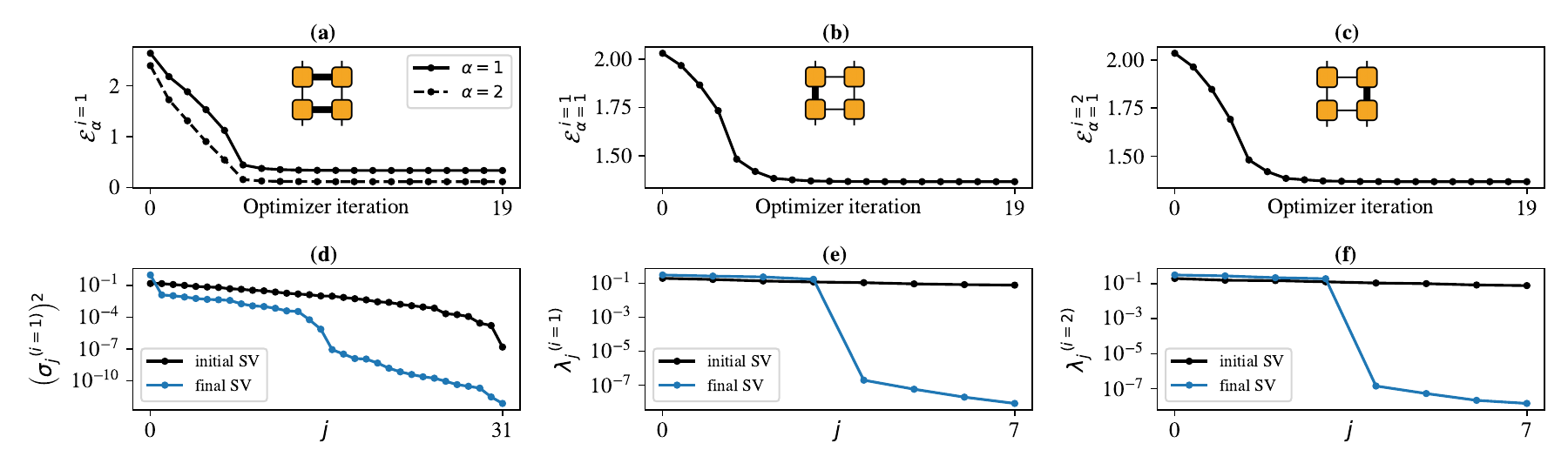}
    \caption{
    Purification optimization on a random state composed of two qudits of dimension $d = 4$.
    For each gradient descent step we show (a) the $\alpha=1$ and $\alpha=2$ entanglement bond entropies, (b) the $\alpha=1$ purification bond entropy on the left site and (c) the $\alpha=1$ purification bond entropy on the right site.
    We plot singular value distributions before and after optimization on (d) the entanglement bond, (e) the left purification bond and (f) the right purification bond. 
    Optimizing only the entanglement bond entropy, we are able to reduce the number of significant singular values on purification bonds down to the physical dimension $d=4$.
    }
    \label{fig:optim_bipartite_state}
\end{figure*}

We highlight the fact that similar techniques for optimizing a purified state entanglement have been introduced for finding the purification entanglement entropy \cite{Nguyen2018} or to compress a \mpdo during unitary or imaginary-time evolution \cite{hauschild2018}.
However, in these works the purification bond dimensions was not changing and was fixed to $2$, eliminating the fundamental issue we are contributing to solve here.
Ref.~\cite{hauschild2018} also employed a different optimization technique to find the disentangler operator, solving a fixed point equation, while we perform a gradient descent.
We believe that gradient descent is a better option, since fixed point iterations are not guaranteed to converge.

The problem of finding a bond factorization that minimizes entanglement was also identified by \cite{werner2016} but applied only to the decomposition of two-qubit gates.
Our method for disentangling two qubits is fairly similar to the technique used in Ref.~\cite{werner2016} to disentangle two-qubit noisy gates.

\section{Results}\label{sec:results}

In this section, we make a detailed numerical analysis of each step of our compression method.
First we focus on the local optimization in a two-qudit case, before exploring the optimization of a multi-qubit \mpdo with sweeps.
Finally, we integrate the technique within a noisy circuit emulator.

\subsection{One-shot compression of a two-qudit \mpdo}
\label{sec:two_qudits}

In Fig.~\ref{fig:optim_bipartite_state} we showcase the compression of a two-qudit \mpdo with the \ourmethod.
We initialize a random \mpdo of two qudits of dimension $d=4$, with (suboptimal) local purification dimensions $r = 8$ and entanglement bond dimension $\chi = 32$.
The constituting tensors were generated with random entries uniformly distributed over the region $[-1, 1) \times [-1, 1)$ of the complex plane.

In panels (a)-(c) we record the variation of the bond entropies against the number of gradient descent steps.
We then optimize the gauge unitary so as to minimize the 2-R\'enyi entropy across the two sites.
As expected, we observe in panel (a) that this procedure decreases the entropy across the entanglement bond; not only the $\alpha=2$ R\'enyi entropy, minimized by construction, but also the $\alpha=1$ entropy. 
As a result, the distribution of singular values becomes much steeper (panel (d)).

Interestingly, this procedure has also the effect of reducing the purification bond entropies (panels (b) and (c)), thereby simplifying the correlations between the physical and ancilla systems.
This translates to a sharp drop in the corresponding (sorted) singular values at index $4$ (panels (e) and (f)), suggesting that the optimization has steered the \mpdo state to a purification of optimal rank.
Indeed, we are now able to truncate the low singular values of index larger than 4 with very high fidelity, thereby reaching a total purification dimension of $4^2$, which is also the dimension of the Hilbert space.
A similar behaviour was observed for other starting values of $d$ and $r$ (not shown).
This demonstrates that it is possible to reduce local purification dimensions through an optimization procedure focused only on the entanglement bonds.

\subsection{One-shot compression of a generic \mpdo}
\label{sec:one_shot_compression}

\begin{figure*}
    \centering
    \begin{subfigure}{\textwidth}
        \centering
        \includegraphics[width=0.96\textwidth]{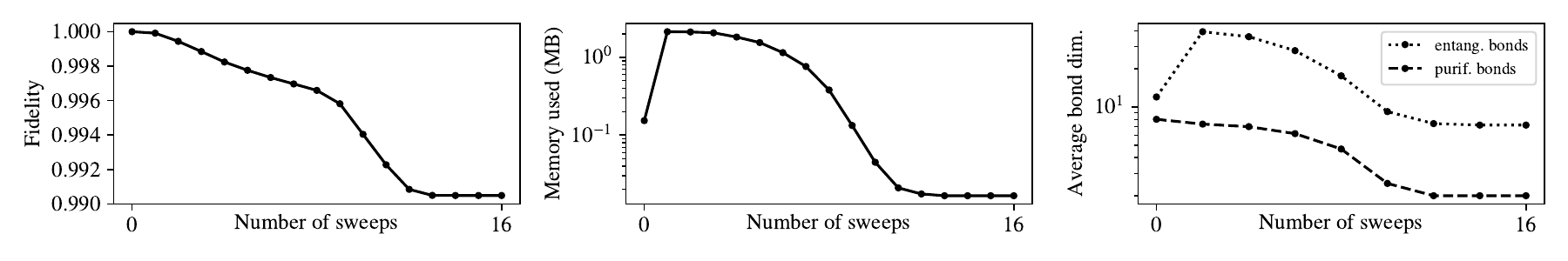}
        \vspace{-4mm}
        \caption{}
        \label{}
    \end{subfigure}
    \begin{subfigure}{\textwidth}
        \centering
        \includegraphics[width=0.96\textwidth]{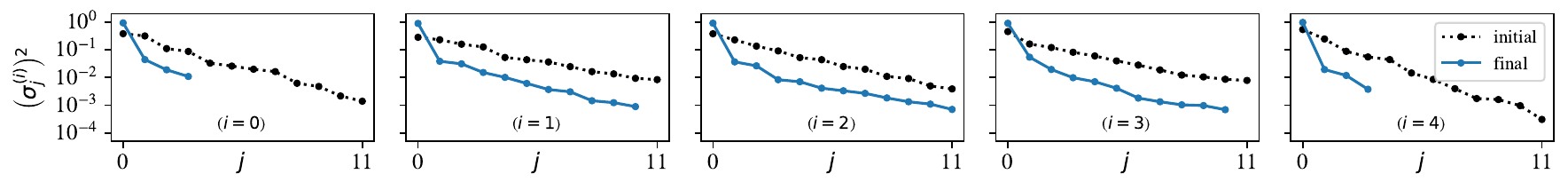}
        \vspace{-2mm}
        \caption{}
        \label{}
    \end{subfigure}
    \begin{subfigure}{\textwidth}
        \centering
        \includegraphics[width=0.96\textwidth]{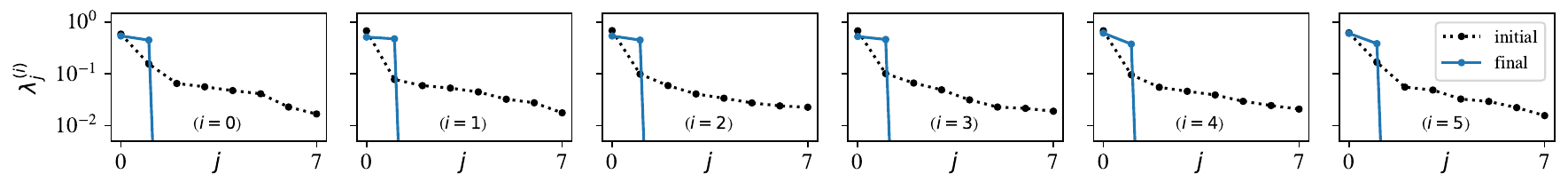}
        \vspace{-2mm}
        \caption{}
        \label{}
    \end{subfigure}
    \caption{
    Local purification disentanglement (16 sweeps, truncation threshold $\varepsilon=10^{-3}$) of a $N=6$ qubits random \mpdo state with initial entanglement bond dimensions $\chi = 12$ and local purification dimensions $r = 8$.
    We show in (a) the variation of total fidelity and memory usage, and the initial and final singular value distributions on (b) entanglement bonds and (c) local purification bond.
    }
    \label{fig:one_shot_compression}
\end{figure*}

Fig.~\ref{fig:one_shot_compression} shows the performance of the \ourmethod on a random \mpdo state of $N=6$ qubits.
Details on the construction of the random state are given in App.~\ref{app:random-state}. We used a noise level $p=10\%$ over ten circuit layers, truncating entanglement bonds to a dimension $\chi=12$ and purification bonds to dimension $r=8$.
We set the truncation threshold to $\varepsilon = 10^{-3}$, close to the smallest singular value on any bond in the starting \mpdo representation.
Fidelities are computed by contracting the tensor networks and using the Uhlmann quantum state fidelity \cite{UHLMANN1976} defined by
\begin{equation}\label{eq:fidelity}
    \mathcal{F}(\rho, \sigma) = \left[ \Tr{\sqrt{\sqrt{\rho} \sigma \sqrt{\rho}}} \right]^2 .
\end{equation}
For the $s$-th sweep, we plot the total fidelity $\mathcal{F}(\rho_s, \rho_0)$, the memory used to store $\rho_s$ in \mpdo form, where $\rho_s$ is the state after the $i$-th disentangling sweep, as well as the average entanglement and purification bond dimensions in the \mpdo representation (panel (a)).
We obtain a gain of an order of magnitude in purification bond dimension and overall memory, while conserving a very good fidelity $> 0.99$.

We observe that, as in the two qudit case, the optimization of local entanglement entropies has a substantial effect on the singular value distributions across local purification bonds.
Indeed, successive sweeps gradually decrease the singular values starting from the third index.
After 16 sweeps these have decreased by several orders of magnitude and can be truncated with high fidelity (panel (c)).
We also observe that singular value distributions on entanglement bonds have become steeper as a result of the local entanglement entropy minimization (panel (b)).
More detailed data including singular value distributions in the intermediate states can be found in App.~\ref{app:convergence}, where we also discuss the convergence of this scheme. 

\begin{figure}[t]
    \centering
    \includegraphics[width=0.48\textwidth]{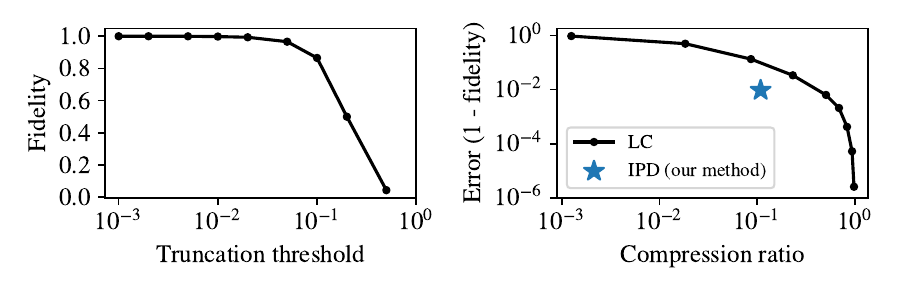}
    \caption{
    Compression results of a truncated SVD scheme applied on the state used in Fig. \ref{fig:one_shot_compression}.
    The truncation threshold corresponds to the ratio of singular values discarded on each bond. The compression ratio is the ratio of memory usage after compression to that before compression.
    }
    \label{fig:naive-one-shot}
\end{figure}

One may ask if this method yields a more precise compression than the \theirmethod, as used in e.g. \cite{white2017} and \cite{cheng2021}.
Indeed our proposed scheme performs a significantly higher number of truncations due to its iterative nature.
In Fig.~\ref{fig:naive-one-shot} we show the compression fidelity and compression ratio obtained by naive compression of the same quantum state as in Fig.~\ref{fig:one_shot_compression} via truncated SVD for different truncation thresholds.
We see that our method is able, for a certain memory compression ratio, to find a better approximation of the initial quantum state; alternatively, for an infidelity budget of $\sim 10^{-2}$, it finds a \mpdo representation an order of magnitude smaller in memory than that found by the \theirmethod. 

This improvement in the one-shot scenario comes with a drawback, which is that the memory usage has to be temporarily increased to accommodate growing entanglement bond dimensions, until further singular values on entanglement bonds fall below the truncation threshold.
In Fig.~\ref{fig:one_shot_compression} one can see that the memory usage increases temporarily by approximately an order of magnitude, before decaying to a value approximately an order of magnitude smaller than its initial size. 
This new compression procedure also experiences an increase in computation time, primarily due to local optimization routines, which become the limiting factor in terms of runtime.

The compression was performed by minimizing the $\alpha=2$ Rényi entropy, as we found that it performed better than minimizing $\alpha=1$.
Not only the gradient is easier to compute (see App.~\ref{app:optimization} for the expressions of each gradient), but the minimization itself converges faster, as described in App.~\ref{app:compare-entropies}.

\subsection{Application to simulating noisy quantum circuits}
\label{sec:application_to_circuits}

\begin{figure*}
    \centering
    \includegraphics[width=0.96\textwidth]{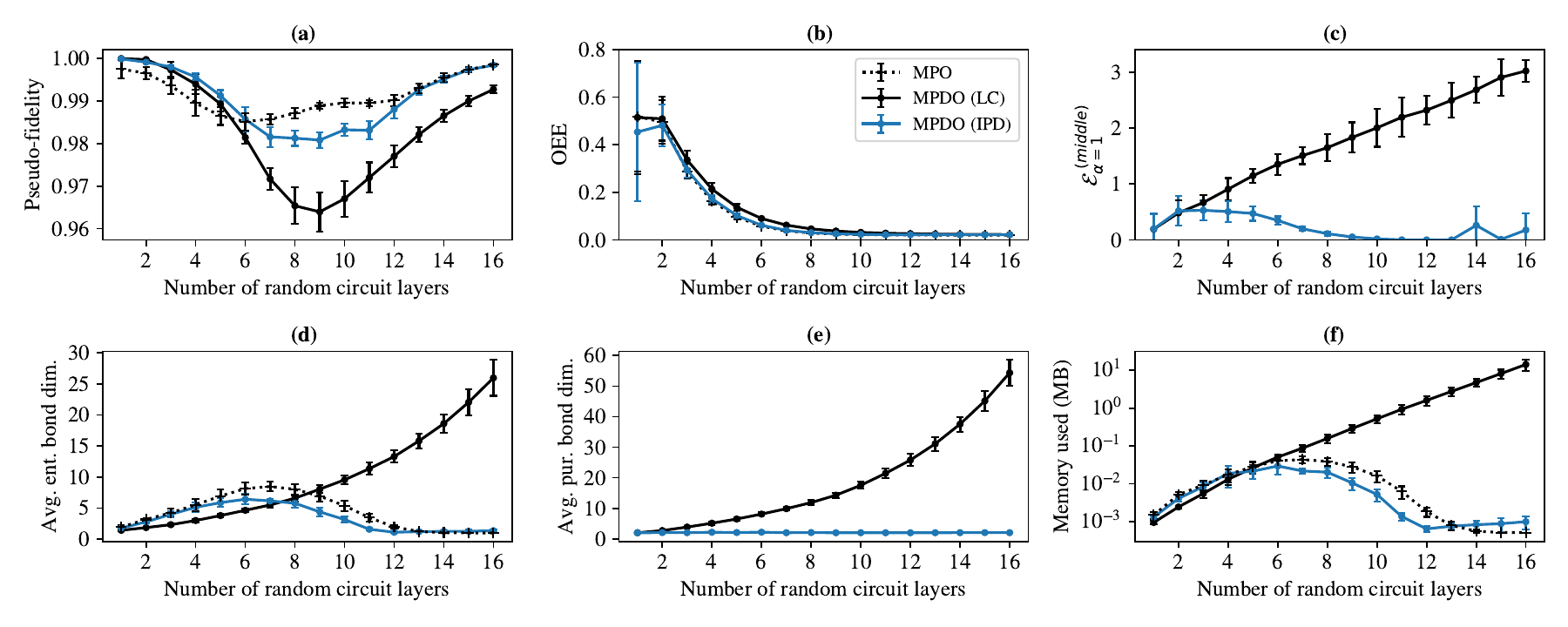}
    \caption{
    Comparison of noisy random circuit ($N=8$ qubits, noise level $p=10\%$) simulations with MPO, \ac{LC}-\mpdo and \ac{IPD}-\mpdo emulators.
    After each circuit layer we show (a) the pseudo-fidelity (defined in Eq.~\eqref{pseudo-fidelity}) with the exact simulation result, (b) the \acf{OEE}, (c) the entanglement entropy of the purified state, (d) the average entanglement bond dimension, (e) the average purification bond dimension and (f) the memory usage after the compression routine.
    Values are averaged over 8 random circuits.
    }
    \label{fig:simu_circuit_with_tol}
\end{figure*}

\begin{figure*}
    \centering
    \includegraphics[width=0.96\textwidth]{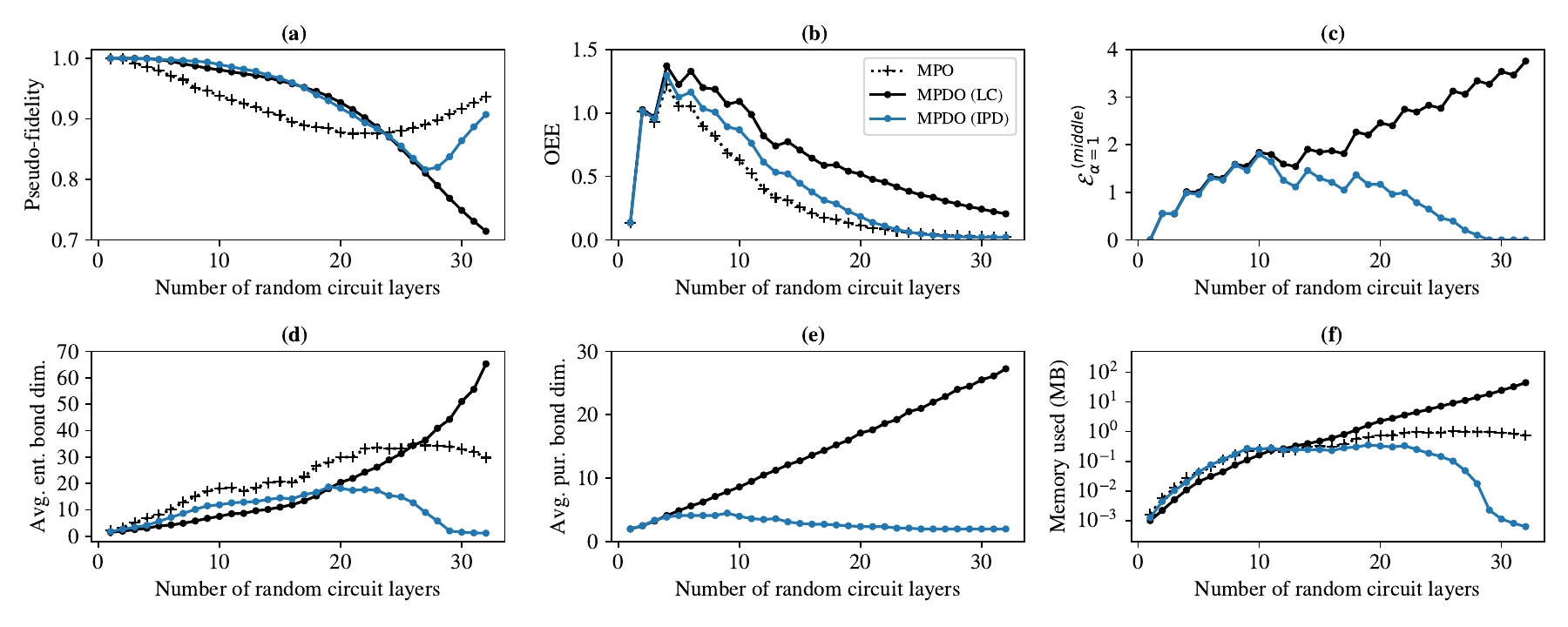}
    \caption{
    Same as Fig.~\ref{fig:simu_circuit_with_tol}, with noise level $p=2\%$. 
    Results for a single random circuit.
    }
    \label{fig:simu_circuit_small_noise}
\end{figure*}

\begin{figure*}
    \centering
    \includegraphics[width=0.96\textwidth]{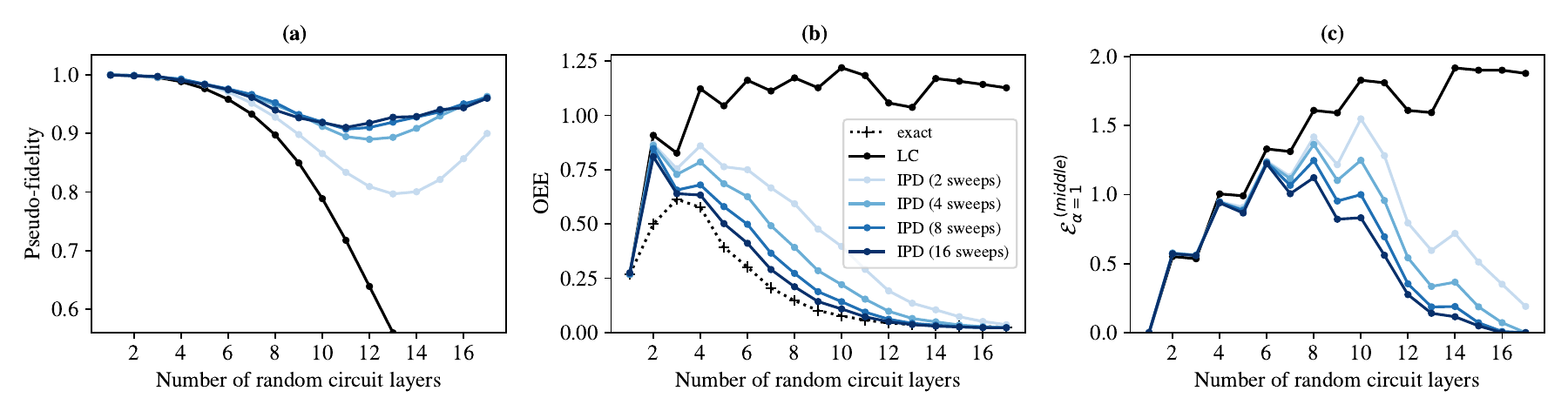}
    \caption{
    Simulation of a noisy random circuit ($N=8$ qubits, noise level $p=5\%$) with fixed bond maximum dimensions and different number of purification disentangling sweeps. 
    After each circuit layer we show (a) the pseudo-fidelity (Eq. \eqref{pseudo-fidelity}) with the exact simulation result, (b) the \acf{OEE} and (c) the entanglement entropy of the purified state.
    }
    \label{fig:simu_circuit_with_bond}
\end{figure*}

We apply the purification disentanglement routine to \mpdo-based noisy circuit simulations to limit the growth in memory usage.
We chose to simulate random circuits sampled as detailed in App.~\ref{app:random-circuit}, with $N=8$ qubits and depolarizing noise
\begin{equation}
    \rho \longrightarrow (1-p)\rho + p \frac{\mathds{1}}{2}
\end{equation}
of various strengths $p$. 
Such a depolarizing noise leads at high depths to a \ac{MMS}, i.e. $\rho_{\rm MMS} = \mathds{1}/2^N$.
Since we want to be able to compress arbitrary states, and not only the trivial \ac{MMS}, we quantified how far we are from it in App.~\ref{app:identity}.

We also compare the results of our simulations with implementations of emulators employing the \theirmethod, in which memory growth is prevented only by performing truncated \acp{SVD} on entanglement and purification bonds. 

In this section we substitute the usual quantum state fidelity with the pseudo-fidelity defined in \cite{WANG2008} by
\begin{equation}\label{pseudo-fidelity}
    \mathcal{F}_\text{alt}(\rho, \sigma) = \frac{\left| \Tr{\rho^\dagger \sigma} \right|}{\sqrt{\Tr{\rho^\dagger \rho} \Tr{\sigma^\dagger \sigma}}}
\end{equation}
for two density operators $\rho$ and $\sigma$.
This is done to accommodate the comparison with MPO simulators, for which Eq.~\eqref{eq:fidelity} may be ill-defined, due to the fact that positive-semidefiniteness is not enforced.

\subsubsection{Simulations with adaptive bond dimensions}

Figure~\ref{fig:simu_circuit_with_tol} shows data for the simulation of a $N=8$ qubit random circuit with noise level $p=10\%$. 
We compare the \ourmethod with MPO and MPDO simulators using the \theirmethod, where truncation is `adaptive', i.e. singular values are truncated up to a certain threshold (tolerance).  
Truncation tolerances for each simulator were tuned so as to obtain comparable memory usage growth at short depths. Values plotted were obtained with $\varepsilon_{MPO} = 0.01$, $\varepsilon_{SC} = 0.015$ and $\varepsilon_{DGC} = 0.001$. For the \ourmethod 8 sweeps were performed after every circuit layer.
A safety maximum bond dimension of 64 (regardless of the bond type) was set for all three simulators. 
The plain blue line corresponds to our method, the plain black line to the \theirmethod, and the dotted black line to an \ac{MPO} simulation.
The data is averaged over $8$ random circuits, with error bars estimated from the standard deviation.

First notice that the \ourmethod performs similarly to \ac{MPO} in terms of fidelity (panel (a)) and memory (panel (f)), while improving largely compared to the \theirmethod.
On panel (a), one can see that the maximal infidelity is half that of the \theirmethod, and is further reduced at large depths $> 13$, while being similar to \ac{MPO} simulation.

Importantly, in our case the memory footprint is bounded, while it grows exponentially with the circuit depth with the \theirmethod.
This is explained by the evolution of both the entanglement and purification bond dimensions, which can be seen averaged over sites respectively in panels (d) and (e).
In the absence of disentanglement, they both increase exponentially with depth, due to the application of entangling gates (for entanglement bonds) and noisy gates (for purification bonds).
With disentanglement, in contrast, the entanglement bond dimensions reach a maximum before going down to one, similarly to the bond dimension of \ac{MPO}; and, strikingly, the purification bond dimensions are reduced down to two, consistently along the circuit at this level of noise.
This value of two is optimal in the sense that one cannot hope to do better for a generic density matrix, as its rank is bounded by the Hilbert space dimension $2^N$.

Panel (b) shows the \ac{OEE} for the three methods. The \ac{OEE} is obtained by contracting the \mpdo states along purification bonds, yielding \ac{MPO} representations, and computing the normalized bond entropy on the middle bond in mixed canonical form. Since \ac{OEE} is a physical quantity, discrepancies are explained by the approximate nature of the simulations, and not by the choice of purification.
Once again, we observe a very good agreement of our method with the \ac{MPO} simulation, better than the \theirmethod.

Finally, panel (c) shows that the entanglement entropy of the purified state, which coincides with the $\alpha=1$ entanglement bond entropy computed on the middle entanglement bond. We observe that while the entropy obtained with the \theirmethod grows in a linear fashion with depth, independently of the \ac{OEE} (understood as a measure of physical entanglement), the \ourmethod allows the simulator to restrain the growth in the entanglement entropy of the purification. This makes explicit the fact that disentangling sweeps are able to find an efficient representation of the state at each depths by finding a purification with low `virtual' entanglement (i.e. low entanglement between ancilla qubits).

Overall, the \ourmethod produces a representation of the state of much better quality than the \theirmethod, and of similar quality as in a \ac{MPO} simulation, with all the advantages of the purification form.
We see in Fig.~\ref{fig:simu_circuit_small_noise} that this conclusion also holds for a weaker noise level $p=2\%$.
In fact, we argue that we improve a bit compared to \ac{MPO}, since we produce, after depth $25$, a representation of significantly smaller memory than \ac{MPO} (see panel (f)), caused by a smaller entanglement bond dimension (panel (d)).

In Fig.~\ref{fig:simu_circuit_small_noise} truncation tolerances for the three emulators were once again tuned so as to produce comparable memory growth at short depths, yielding $\varepsilon_{MPO} = 0.015$, $\varepsilon_{SC} = 0.005$ and $\varepsilon_{DGC} = 0.0005$. More significantly, the number of sweeps performed after every circuit layer was increased to 32.

Another difference with the stronger noise case is that the purification bond dimensions (panel (e)) show a maximum ($< 5$) before being reduced to the optimal value of two.
A lower noise level indeed leads to less entanglement destruction, and thus an entanglement pattern that is more complex to disentangle for our algorithm.

\subsubsection{Performance at fixed bond dimensions}

Thankfully, in our approach one can increase the number of sweeps to improve the quality of the optimization, and thus of the purification.
This is illustrated in Fig.~\ref{fig:simu_circuit_with_bond}, which compares the pseudo-fidelity, the \ac{OEE} and the entanglement entropy of the purified state (the latter two computed on the middle bond) with an increasing number of disentangling sweeps per circuit layer.
Here, to make a fair comparison, we avoid the somewhat arbitrary nature of the truncation tolerance, by fixing all bond dimensions of the \mpdo and focus on entropies rather than bond dimensions. Entanglement bonds are truncated to a maximum dimension of $\chi=32$, while purification bonds are truncated to the rank-optimal dimension $r=d=2$. This truncation is performed after every circuit layer (and after the disentangling sweeps in the case of the \ourmethod).
The circuit simulated is a single random quantum circuit with $8$ qubits as used in the results of Fig.~\ref{fig:simu_circuit_with_tol}. We used a depolarizing noise level of $p=5\%$.
We clearly see an improvement in fidelity (panel (a)) and a systematic decrease in \ac{OEE} (panel (b)) and purification entropy (panel (c)) with more sweeps.
An decrease in entropy translates into a reduction in memory, as it directly shows that less singular values need to be kept.
Note how the \ac{OEE} gets closer to the \ac{MPO} simulation values (panel (b)), which seems to act as lower bound.

These results indicate that, with a correct disentangling technique such as the one we propose, the representative power of \mpdo is as good as \ac{MPO} all along the circuit.
In their recent work~\cite{guo_locally_2023}, Guo and Yang argue on the contrary that \mpdos have a good representative power only at low depths and at high depths, and lack representative power in between when compared to \acp{MPO}.
They also observe that the separation between these two regimes is close to the maximum of \ac{OEE}.
If we indeed observe a minimum of fidelity and a maximum of memory with our \mpdos at intermediate depths, the values are still fairly close to the \ac{MPO} simulation, and thus reflect an increase of difficulty that is shared with \ac{MPO} techniques.
In addition, we see no co-location of this difficulty barrier with the maximum of \ac{OEE}.
For example in Fig.~\ref{fig:simu_circuit_with_tol}, the maximum of \ac{OEE} lies around depth $3$, while the minimum of fidelity moves between $10$ and $14$ depending on the number of sweeps, and the maximum of purification entropy is between depths $6$ and $10$.

This indicates that the barrier observed in the data of Ref.~\cite{guo_locally_2023} is caused by the \mpdo compression they perform and by the \ac{MPO}-to-\mpdo mapping they use.
The \ourmethod, although more involved computationaly, leads to \mpdos of better quality (higher fidelity, lower memory), and are thus able to overcome the complexity barrier of Ref.~\cite{guo_locally_2023}.
One can hypothesize that we explore a larger subspace of equivalent purifications, which allows for a more compact representation of the state.

\section{Discussion and conclusion}
\label{sec:conclusion}

In conclusion, we presented a compression technique for \mpdos exploiting the freedom of choosing the basis for the environment qudits in a purification.
Instead of optimizing this choice locally on each qubit, we perform it globally, by applying optimized changes of basis on successive pairs of qudits, organized in sweeps. 
This amounts to disentangling the purified state in view of finding a representation with low entanglement and purification bond entropies.
We then truncate the smallest singular values to finish the compression.

We find that a good criterion for this local optimization is to minimize the entanglement bond entropy.
With this choice of cost function, we observe, surprisingly, a reduction of \emph{both} the entanglement and purification entropy.
This is a surprisingly simple cost function for the minimization of two seemingly distinct entanglements sources.
We recall that entanglement bond entropy is produced by entangling gates, while purification bond entropy is produced by noisy gates.
If a relationship exists between these, it is not obvious with \mpdos.
For instance, noisy gates, when they are a source of decoherence, usually destroy entanglement\cite{noh2020}, but this does not automatically translate,  in \mpdos, to the reduction of entanglement bond entropies.

In a way, one can see our compression scheme as a way to recover this relationship.
Another way to interpret this choice of cost function is that the \ourmethod is similar in spirit as trying to find the entanglement of purification~\cite{terhal2002}, which is in our language the minimal entropy of a given entanglement bond, over all equivalent purifications.
This quantity captures the overall complexity of a bipartite mixed state, including both quantum and classical correlations~\cite{terhal2002}.
As classical correlations come from noisy gates, it seems natural in this context that reducing entanglement bond entropy leads also to a reduced purification bond entropy.

In fact, for every state we explored, we were able to reduce the purification bond dimensions to a value close to $2$, with a very good truncation fidelity.
This value of $2$ is optimal for a generic state, in the sense that it corresponds to a decomposition $\rho = X^\dag X$ with $X$ of dimension $2^N \times 2^N$, which is minimal given a generic $\rho$ with rank $2^N$.
It is left to future work to establish if our technique can reduce $r$ below this value when the density matrix is of lower rank, such as when noise is low.

We also compared two measures of entanglement in App.~\ref{app:compare-entropies} and found that locally minimizing the 2-R\'enyi entanglement bond entropy yields a better final purification with faster convergence.

This work also brings a new perspective on the question of the representability of practical states with \mpdos. 
The existence of states that are easily represented as \acp{MPO} but difficult as \mpdos has been proved in Ref.~\cite{cuevas2013}.
However, this does not close the debate for practical states, i.e. states encountered in typical quantum computing programs.
Our results seem to indicate that \mpdos have a similar representation power compared to \acp{MPO}, at least for noisy random circuits.
These contradict the conclusions of Ref.~\cite{guo_locally_2023}, which observed a sharp loss of representability close to the maximum of \ac{OEE}, but used simpler compression methods. 
Using the \ourmethod, it seems that a good \mpdo representation of practical states can always be found.
It is possible however, that such a good representation is harder to reach than for \acp{MPO}.
This would not be surprising given the advantages of \mpdos over \acp{MPO}: enforcing positive-semidefiniteness, and providing efficient computation of observables, trace and reduced density matrices thanks to its canonical form.
Further work is needed to sharpen this point.

Despite being efficient at compressing a \mpdo, the \ourmethod in its present version is not computationally cheap.
Further studies are required to better understand how the compression ratio scales with the number of qubits, and how it is impacted by 2D geometry, in order to evaluate if it is a good option for the simulation of circuits of practical sizes and connectivities.
In addition, the tuning of parameters (number of sweeps, truncation thresholds, etc) could be optimized depending on noise level.
Performances would also benefit from improvements of the optimization under unitary constraints.
For all these reasons, the precise assessment of computation times is left for further works.

\acknowledgments

This work is part of HQI initiative (\url{www.hqi.fr}) and is supported by France 2030 under the French National Research Agency award number “ANR-22-PNCQ-0002”.
Computations have been done on Eviden Qaptiva.

\bibliographystyle{apsrev4-2}
\bibliography{biblio}

\clearpage

\appendix 

\setcounter{figure}{0}
\renewcommand\thefigure{A.\arabic{figure}}

\section{Derivation of the gradient for gauge optimizations}\label{app:optimization}

In this section we compute the gradient used to solve the optimization problem in Eq.~\eqref{eq:entropy-opt}. We adopt the notation of Fig.~\ref{fig:entglt_optim} where $A$ denotes the contraction of two adjacent sites in mixed canonical form and $\tilde A$ is the contraction of $A$ with the gauge unitary $U$.

\begin{figure}[t]
	\centering
	\includegraphics[width=0.3\textwidth]{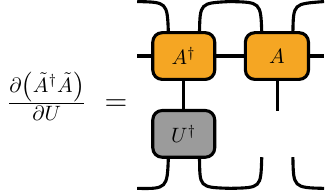}
	\caption{Tensor network expression of the partial derivative of $X = \tilde A^\dagger \tilde A$ with respect to the unitary $U$.}
	\label{fig:dM_dU}
\end{figure}

Letting $X= \tilde A ^\dagger \tilde A$, the chain rule reads 
\begin{equation}
    \frac{\partial E}{\partial U} = \sum_{ij}  \frac{\partial X_{ij}}{\partial U} \frac{\partial E}{\partial X_{ij}} , 
\end{equation}
where $\frac{\partial X_{ij}}{\partial U}$ can be obtained via the tensor expression in Fig.~\ref{fig:dM_dU}. Below we compute $\frac{\partial E}{\partial U}$ for the R\'enyi entropies in Eq.~\eqref{renyi-entropies} with $\alpha=1$ and $\alpha=2$.

\subsection{R\'enyi entropy with $\alpha=1$}

The 1-R\'enyi entropy is given by
\begin{equation}
    E(X) = - \Tr{X \ln X } .
\end{equation}
Differentiating with respect to elements of $X$ yields
\begin{equation}\label{eq:dE_dM}
    \frac{\partial E(X)}{\partial X_{ij}} = - \Tr{\frac{\partial X}{\partial X_{ij}} \ln X + X \frac{\partial \ln X}{\partial X_{ij}} } .
\end{equation}
The first term is straightforward to simplify, since
\begin{equation}
    \left( \frac{\partial X}{\partial X_{ij}} \right)_{k\ell} = \delta_{ik}\delta_{j\ell} .
\end{equation}
One then obtains
\begin{align}
    \Tr{\frac{\partial X}{\partial X_{ij}} \ln X} &= \sum_k \sum_\ell \left( \frac{\partial X}{\partial X_{ij}} \right)_{k\ell} (\ln X)_{\ell k} = (\ln X)_{ji} .
\end{align}
We now turn to the second term of (\ref{eq:dE_dM}). For a positive definite matrix $M$ of unit trace, the logarithm is well defined and the series
\begin{equation}
    \ln X = \sum_{n=1}^\infty \frac{(-1)^{n-1}}{n} (X - \mathds{1})^n 
\end{equation}
converges since $\| X-\mathds{1} \|_\infty < 1$ . By the product rule we have
\begin{equation}
    \frac{\partial}{\partial X_{ij}} (X-\mathds{1})^n = \sum_{k=0}^{n-1} (X-\mathds{1})^k \frac{\partial X}{\partial X_{ij}} (X-\mathds{1})^{n-1-k} ,
\end{equation}
and using the cyclic property of the trace, we obtain
\begin{equation}
    \Tr{X \frac{\partial \ln X}{\partial X_{ij}}} = \Tr{ X \frac{\partial X}{\partial X_{ij}} \sum_{n=0}^\infty (\mathds{1}-X)^n } .
\end{equation}
Moreover, since $X = \mathds{1} - (\mathds{1} - X)$, we obtain a telescopic sum, so that only the $n=0$ term survives, and
\begin{equation}
    \Tr{X \frac{\partial \ln X}{\partial X_{ij}}} = \Tr{\frac{\partial X}{\partial X_{ij}}} = \delta_{ij} .
\end{equation}
Finally,
\begin{equation}
    \frac{\partial E(X)}{\partial X} = - \left[(\ln X)^T + \mathds{1}\right] .
\end{equation}

\begin{figure*}
	\centering
	\begin{subfigure}{\textwidth}
		\centering
		\includegraphics[width=0.96\textwidth]{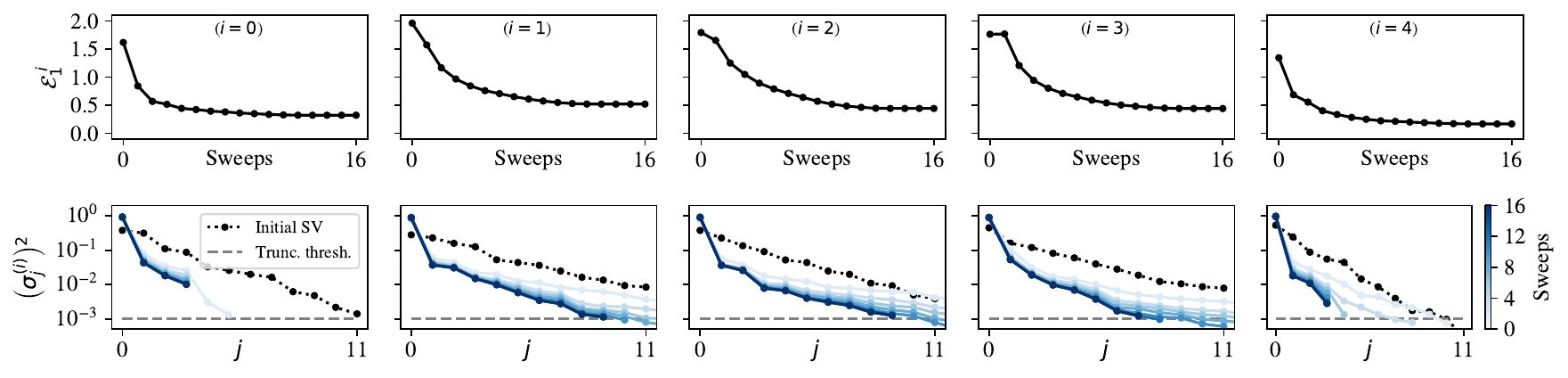}
		\vspace{-2mm}
		\caption{}
		\label{}
	\end{subfigure}
	\begin{subfigure}{\textwidth}
		\centering
		\includegraphics[width=0.96\textwidth]{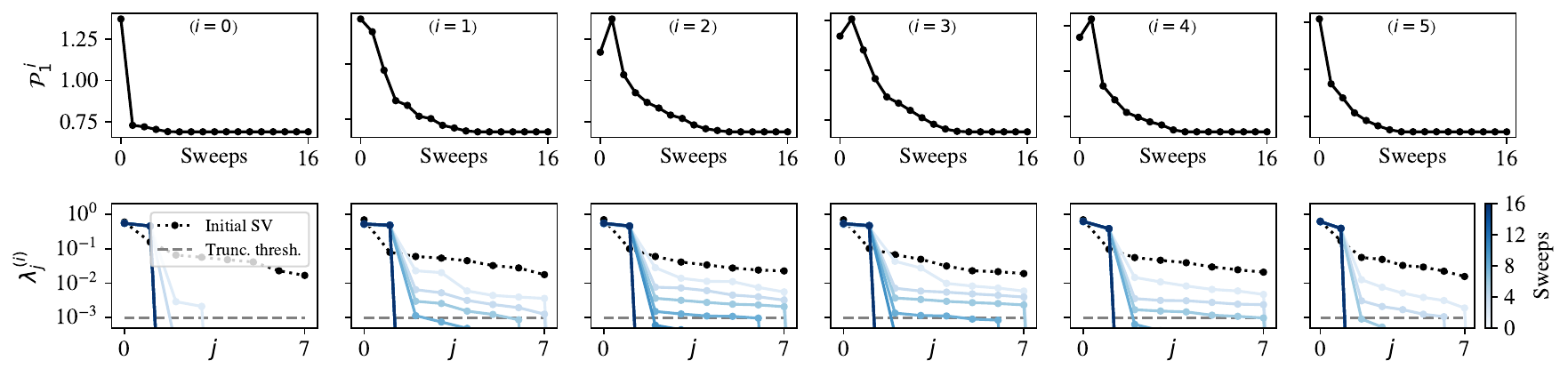}
		\vspace{-2mm}
		\caption{}
		\label{}
	\end{subfigure}
	\caption{
		Convergence analysis for the one-shot purification disentanglement of a random \mpdo shown in Fig.~\ref{fig:one_shot_compression}.
		The convergence of local bond entropies and the variation of singular value distributions on each bond over the number of sweeps are shown for entanglement bonds in panel (a) and purity bonds in panel (b).
	}
	\label{fig:convergence}
\end{figure*}

\subsection{R\'enyi entropy with $\alpha=2$}

We now compute the partial derivative of the 2-R\'enyi entropy
\begin{equation}\label{renyi-entropy}
    E_2(X) = - \ln\Tr{X^2}.
\end{equation}
with respect to elements of $X$. First we note that
\begin{equation}
    \frac{\partial}{\partial X_{ij}} \left( -\ln \Tr{X^2} \right) = -\frac{1}{\Tr{X^2}} \Tr{\frac{\partial X^2}{\partial X_{ij}}} .
\end{equation}
Then, by cyclicity
\begin{equation}
    \Tr{\frac{\partial X^2}{\partial X_{ij}}} = 2 \Tr{\frac{\partial X}{\partial X_{ij}}X} ,
\end{equation}
and
\begin{equation}
    \Tr{\frac{\partial X}{\partial X_{ij}}X} = \sum_{k\ell} \frac{\partial X_{k\ell}}{\partial X_{ij}} X_{\ell k} = \sum_{k\ell} \delta_{ik} \delta_{j\ell} X_{\ell k} = X_{ji},
\end{equation}
from which
\begin{equation}
    \frac{\partial E_2(X)}{\partial X} = - \frac{2 X^T}{\Tr{X^2}}.
\end{equation}

\section{Details of convergence in the one-shot scenario}\label{app:convergence}

Here we study in more detail the optimization performed in Fig.~\ref{fig:one_shot_compression} and discuss convergence of the \ourmethod.

\begin{figure}[h]
    \centering
    \includegraphics[width=0.48\textwidth]{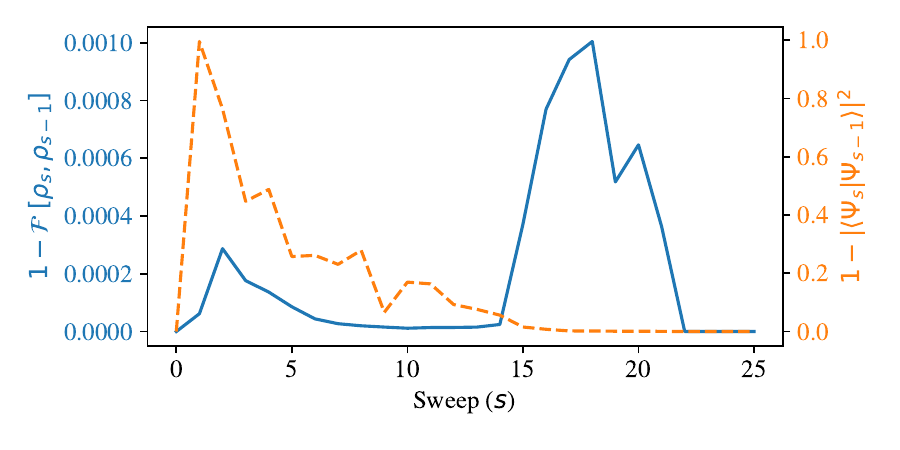}
    \caption{
    Convergence study of the procedure of Fig.~\ref{fig:one_shot_compression}.
    We show the infidelity of the $s$-th sweep for the state (plain blue line, left-hand-side scale) and for its purification (dashed orange line, right-hand-side scale).
    $\ket{\Psi_{s}}$ and $\rho_s$ denote respectively the purification and the state obtained after the $s$-th sweep.
    }
    \label{fig:relative_fidelities}
\end{figure}

Fig.~\ref{fig:convergence} shows a more complete optimization report of the one-shot purification disentanglement procedure performed in Fig.~\ref{fig:one_shot_compression}.
We show the variation of bond entropies (of order $\alpha=1$) $\mathcal{E}_1^i$ and $\mathcal{P}_1^i$ (defined in Eq.~\eqref{eq:bond-entropies}) with the number of sweeps $s$, and the bond singular value distributions $\{(\sigma_j^{(i)})^2\}_j$ and $\{\lambda_j^{(i)}\}_j$ after different numbers of sweeps.
We observe that bond entropies, for both entanglement bonds and purification bonds, decrease with the number of sweeps and seem to converge to non-zero values. 

One would be justified in asking whether this iterative disentanglement technique converges to one final purified state.
To answer this we repeat the computations reported in Fig.~\ref{fig:one_shot_compression}, this time without truncating purification bonds between sweeps.
This allows us to keep the dimension of the ancilla system constant and study the variation of the purified state.
In Fig.~\ref{fig:relative_fidelities} we observe that the distance between purifications obtained after successive sweeps, quantified using the fidelity, reliably decreases to zero (dashed orange line).
After about 15-20 sweeps, the local disentangling operations converge in one or a few gradient descent steps, and the procedure has thus converged to a locally optimized purification.

Fig.~\ref{fig:relative_fidelities} also records the relative error of the state due to the necessary truncation of entanglement bonds.
The behavior for the first 10-15 sweeps is similar to the purification relative distance measure, and can be explained by the fact that early sweeps tend to increase entanglement bond dimensions significantly, yielding a non-negligible error upon truncation.
Moreover, we observe a peak in the value of $1-\mathcal{F}[\rho_s, \rho_{s-1}]$ in the range of 15-20 sweeps.
This can be identified in Fig.~\ref{fig:convergence} with the region in which a significant number of singular values on entanglement bonds fall below the truncation threshold, thus accounting for most of the truncation error.

\begin{figure}
    \centering
    \includegraphics[width=0.48\textwidth]{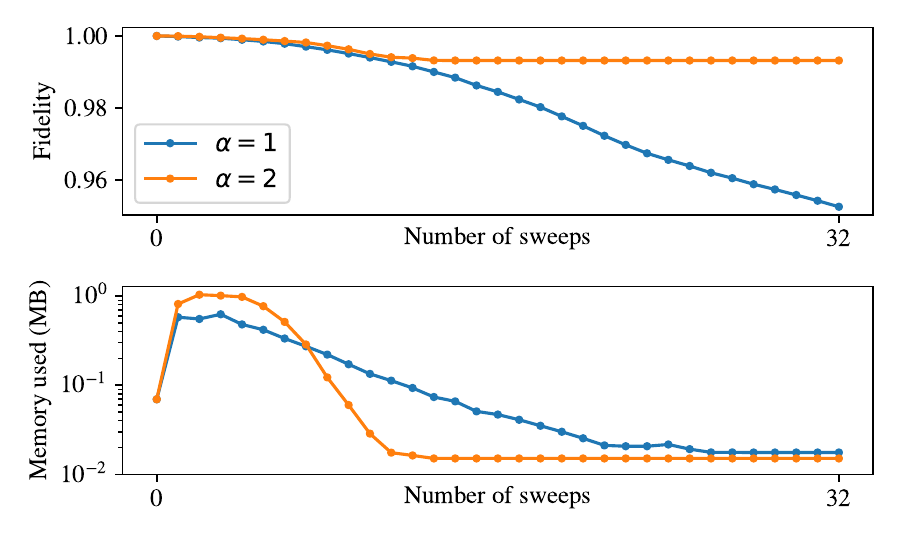}
    \caption{Comparison of $\alpha=1$ and $\alpha = 2$ cost functions in the one-shot disentanglement scenario. We plot the exact compression fidelity (top) and the memory usage (bottom) in both cases.}
    \label{fig:comparison_renyi}
\end{figure}

\section{Comparison of performance with different entropy metrics}\label{app:compare-entropies}

In this section we justify numerically our choice of using the $\alpha=2$ entanglement bond entropy as the objective function in Eq.~\eqref{eq:entropy-opt}, as opposed to $\alpha=1$.

In Fig.~\ref{fig:comparison_renyi} we compare the results of the one-shot disentanglement compression of a random quantum state, using as an objective function the $\alpha=1$ (blue line) and $\alpha=2$ (orange line) entanglement bond entropies.  

The quantum state is generated by simulating 6 layers of a 6-qubit random circuit with depolarizing noise level $p=5\%$.
At each layer both entanglement bonds and purifications bond are held fixed by truncated SVD, at dimensions $\chi=r=8$.
In the purification disentanglement sweeps we used a truncation threshold of $\varepsilon = 10^{-3}$ and a maximum allowed bond dimension $\chi_{max} = 64$.
Local optimizations were implemented with a conjugate gradient algorithm with tolerance $10^{-5}$ and a maximum of 500 iterations.

We observe that locally minimizing the $\alpha=2$ entropies allow one to converge to a low-memory solution faster and with much better fidelity that minimizing $\alpha=1$ entropies.
A closer study of singular value distributions across sweeps (not shown) reveals that the $\alpha=1$ entropy optimization results in a faster decay of entanglement bond dimensions, but does not exhibit the ``plateau descent'' of singular values on purification bonds observed in Fig.~\ref{fig:convergence} with $\alpha=2$.
A similar behavior was observed for other parameters of the quantum state generation or the disentangling procedure, which also translated to generally better performances for $\alpha=2$ optimizations in circuit simulations.

\section{Random noisy circuits}
\label{app:random-circuit}
\label{app:random-state}

\begin{figure}[h]
    \centering
    \includegraphics[width=0.24\textwidth]{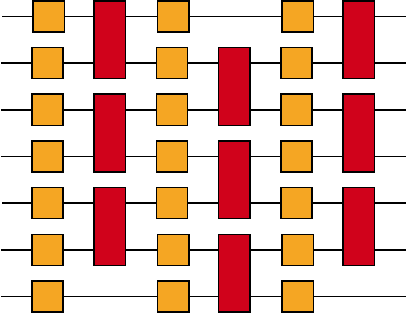}
    \caption{Illustration of three layers of a random quantum circuit as used in \cite{cheng2021}.}
    \label{fig:random_circuit}
\end{figure}

To benchmark our circuit emulators we use random quantum circuits.
We consider similar circuits as in Ref.~\cite{cheng2021}.
A circuit layer consists of a layer of one-qubit gates followed by a layer of nearest-neighbor (in a 1D topology) two-qubit gates, alternating as illustrated in Fig.~\ref{fig:random_circuit}.
Such circuits were also used for generating random \mpdos in Sec.~\ref{sec:one_shot_compression} and App.~\ref{app:convergence} and~\ref{app:compare-entropies}, with truncation to fixed bond dimensions after each circuit layer. 

\begin{figure*}[t]
    \centering
    \includegraphics[width=0.96\textwidth]{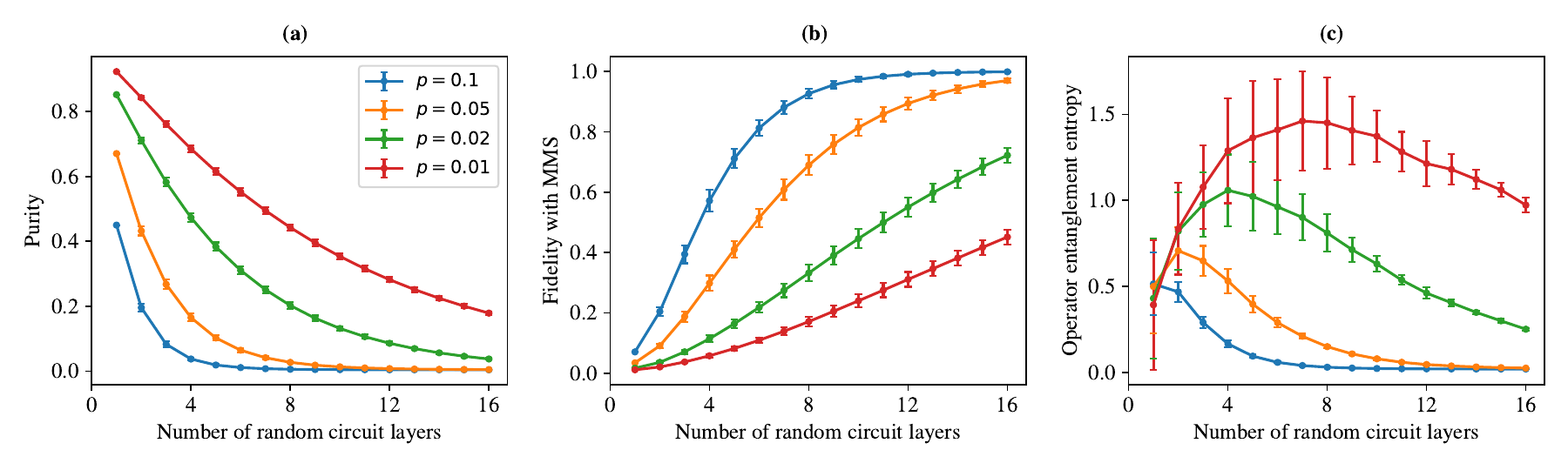}
    \caption{
    Convergence to a \acf{MMS} at large depths for $N=8$ qubits random circuits.
    This is quantified at various depolarizing noise levels $p$ with three quantities: purity (a), fidelity to the \ac{MMS} (b) and \ac{OEE} (c).
    Values are averaged over 16 random circuits and error bars show the corresponding standard deviation.
    }
    \label{fig:identity}
\end{figure*}

The one-qubit gates are parametrized rotations defined by the unitary matrix $\exp{i\alpha (\sigma_x \sin\theta \cos\phi + \sigma_y \sin\theta\sin\phi + \sigma_z \cos\theta)}$, where the parameters $\alpha$, $\theta$, $\phi$ are chosen uniformly at random in $[0, 2\pi)$ for each gate, and $\sigma_x, \sigma_y, \sigma_z$ are the three Pauli matrices.
The two-qubit gates are chosen to be either controlled-NOT (CNOT) or controlled-Z (CZ) gates with equal probability.
A depolarizing noise channel,
\begin{equation}\label{eq:noise}
    \rho \longrightarrow (1-p)\rho + p \frac{\mathds{1}}{2} ,
\end{equation}
with noise level $p$ is applied qubit-wise after each layer.

\section{Proximity to the \acf{MMS}}
\label{app:identity}

As depolarizing noise leads ultimately to the \acf{MMS}, i.e. $\rho_{\rm MMS} = \mathds{1}/2^N$, we want to understand if the performances of our compression is caused by the proximity to this state.
Indeed, the \ac{MMS} is a trivial state to represent with an \mpdo of purification bond dimensions $r_i = 2$ and entanglement bond dimensions $\chi_i = 1$.
One could question if the reduction of purification bonds we obtain is simply a demonstration that a \ac{MMS} has been reached.

In Fig.~\ref{fig:identity}, we quantify the proximity to the \ac{MMS} along a noisy random circuit through three quantities: purity, fidelity against the \ac{MMS} and \ac{OEE}.
For the \ac{MMS}, these should be zero, one and zero respectively.
Exact quantum states for each depth are computed and subsequently mapped to a MPO via successive \acp{SVD} to measure the \ac{OEE}.

For noise level $p=10\%$ (blue line), about ten layers are needed to reach the \ac{MMS}.
However, we see in Fig~\ref{fig:simu_circuit_with_tol} that the \ac{IPD} yields purification bond dimensions $r_i \approx 2$ even for the 10 first layers.
This means the compression of the purification bond dimensions was obtained for states that are far from the \ac{MMS}, with substantial purity and \ac{OEE}.

Performing the same analysis with the $p=2\%$ case (green line), and comparing to Fig.~\ref{fig:simu_circuit_small_noise}, gives the same conclusion.
This time, purification bond dimensions $r_i \approx 2$ is reached after about 15 layers, at which the \ac{OEE} and infidelity to the \ac{MMS} are not negligible.
We note that during the first layers, where purification bond dimensions are a bit higher ($r_i \le 5$), purity is also larger.
This suggests a relation between optimized purification bond dimension and purity.

Note also that the random state used in App.~\ref{app:compare-entropies} has purity $\approx 0.1$ and fidelity $\approx 0.5$ with respect to the \ac{MMS}.

\section{List of naming conventions for MPOs and MPDOs}
\label{app:terminology}

The literature about MPDOs and MPOs produced a surprising amount of different and sometimes mutually exclusive naming conventions.
To help the new reader navigate through them, we compiled a list of those we encountered in Table~\ref{tab:mpdo_lexicon}.

\begin{table*}[b]
    \centering
    \begin{tabular}{|c|c|c|}
\hline 
Reference & name for MPDO & name for MPO
\tabularnewline
\hline 
\hline 
\citet{cheng2021} & MPDO &\tabularnewline
\hline 
\citet{DeLasCuevas2020} & LPF\footnote{Locally purified form, not an acronym in the text.} & MPDO / MPO \tabularnewline
\hline 
\citet{guo2022},~\citet{surace2019} & MPDO &  MPO \tabularnewline
\hline 
\citet{guo_scalable_2023},~\citet{guo_locally_2023} & LPDO &  MPO \tabularnewline
\hline 
\citet{werner2016} & LPTN & MPO \tabularnewline
\hline 
\citet{jaschke2018} & LPTN & MPDO \tabularnewline
\hline 
\citet{hauschild2018} & purification & MPO \tabularnewline
\hline 
\citet{cuevas2013} & local purification &  MPDO \tabularnewline
\hline 
\citet{Verstraete2004} & MPDO & MPDO\tabularnewline
\hline 
\citet{Prosen2009},~\citet{Oh2021},~\citet{noh2020} & & MPO\tabularnewline
\hline 
\citet{zwolak_mixed_state_2004} & & MPD\tabularnewline
\hline 
\citet{white2017},~\citet{GuthJarkovsky2020} & & MPDO\tabularnewline
\hline 
\end{tabular}
    \caption{Guide for the reader of the MPDO and MPO literature.}
    \label{tab:mpdo_lexicon}
\end{table*}

\end{document}